# Single shot amplitude and phase characterization of optical arbitrary waveforms

V.R. Supradeepa, Daniel E. Leaird and Andrew M. Weiner

*School of Electrical and Computer Engineering, Purdue University, West Lafayette, Indiana 47907, USA.*

*venkatas@purdue.edu, leaird@purdue.edu, amw@ecn.purdue.edu.*

**Abstract**: Using a time-gated dual quadrature spectral interferometry technique, for the first time we demonstrate single-shot characterization of both spectral amplitude and phase of ~1THz bandwidth optical arbitrary waveforms generated from a 10 GHz frequency comb. Our measurements provide a temporal resolution of 1ps over a record length of 100ps. Single-shot characterization becomes particularly relevant when waveform synthesis operations are updated at the repetition rate of the comb allowing creation of potentially infinite record length waveforms. We first demonstrate unambiguous single shot retrieval using rapidly updating waveforms. We then perform additional single-shot measurements of static user-defined waveforms generated via line-by-line pulse shaping.

**References and links**

1. Z. Jiang, D. S. Seo, D. E. Leaird and A. M. Weiner. "Spectral line by line pulse shaping," Opt. Lett. **30**, 1557–1559 (2005).
2. Z Jiang, CB Huang, DE Leaird, AM Weiner, "Optical arbitrary waveform processing of more than 100 spectral comb lines," Nature Photonics **1**, 463-467 (2007).
3. R. P. Scott, N. K. Fontaine, J. Cao, K. Okamoto, B. H. Kolner, J. P. Heritage, and S. J. B. Yoo, "High-fidelity line-by-line optical waveform generation and complete characterization using FROG," Opt. Express **15**, 9977-9988 (2007).
4. D. Miyamoto; K. Mandai; T. Kurokawa; S. Takeda; T. Shioda; H. Tsuda, "Waveform-Controllable Optical Pulse Generation Using an Optical Pulse Synthesizer," IEEE Photon. Technol. Lett. **18**, 721–723 (2006).
5. K. Takiguchi, K. Okamoto, T. Kominato, H. Takahashi, and T. Shibata, "Flexible pulse waveform generation using silica-waveguide-based spectrum synthesis circuit," Electron. Lett. **40**, 537–538 (2004).
6. V. R. Supradeepa, Chen-Bin Huang, Daniel E. Leaird, and Andrew M. Weiner, "Femtosecond pulse shaping in two dimensions: Towards higher complexity optical waveforms," Opt. Express **16**, 11878-11887 (2008).
7. R. M. Huffaker, R. M. Hardesty, "Remote sensing of atmospheric wind velocities using solid-state and $CO_2$ coherent laser systems," Proceedings of the IEEE **84**, 181-204 (1996).
8. W. S. Warren, H. Rabitz, and M. Dahleh, "Coherent control of quantum dynamics:the dream is alive," Science **259**, 1581 (1993).
9. J. T. Willits, A. M. Weiner, and S. T. Cundiff, "Theory of rapid-update line-by-line pulse shaping," Opt. Express **16**, 315-327 (2008).
10. D. J. Kane; R. Trebino, "Characterization of arbitrary femtosecond pulses using frequency-resolved optical gating," IEEE J. Quantum Electron. **29**, 571-579 (1993).
11. P. O'Shea, M. Kimmel, X. Gu, and R. Trebino, "Highly simplified device for ultrashort-pulse measurement," Opt. Lett. **26**, 932-934 (2001).
12. C. Iaconis and I. A. Walmsley, "Spectral phase interferometry for direct electric-field reconstruction of ultrashort optical pulses," Opt. Lett. **23**, 792-794 (1998).
13. V. R. Supradeepa, Daniel E. Leaird, and Andrew M. Weiner, "Optical arbitrary waveform characterization via dual-quadrature spectral interferometry," Opt. Express **17**, 25-33 (2009).
14. J. Bromage, C. Dorrer, I. A. Begishev, N. G. Usechak, and J. D. Zuegel, "Highly sensitive, single-shot characterization for pulse widths from 0.4 to 85 ps using electro-optic shearing interferometry," Opt. Lett. **31**, 3523-3525 (2006).

15. Houxun Miao, Daniel E. Leaird, Carsten Langrock, Martin M. Fejer, and Andrew M. Weiner, "Optical arbitrary waveform characterization via dual-quadrature spectral shearing interferometry," Opt. Express **17**, 3381-3389 (2009)
16. M. A. Foster, R. Salem, D. F. Geraghty, A. C. Turner-Foster, M. Lipson, and A. L. Gaeta, "Silicon-chip based ultrafast optical oscilloscope," Nature **456**, 81-84 (2008).
17. C. V. Bennett, B. D. Moran, C. Langrock, M. M. Fejer, and M. Ibsen, "640 GHz real time recording using temporal imaging." CtuA6, Conference on Lasers and Electro-Optics (CLEO) (2008)
18. John E. Heebner, Chris H. Sarantos, "Progress Towards the Solid-State All-Optical Streak Camera", CThW1, Conference on Lasers and Electrooptics (CLEO) (2009).
19. C. Dorrer, "High-speed measurements for optical telecommunication systems," IEEE J. Sel. Top. Quantum Electron. **12**, 843-858, (2006).
20. L. Lepetit, G. Cheriaux, and M. Joffre, "Linear techniques of phase measurement by femtosecond spectral interferometry for applications in spectroscopy," J. Opt. Soc. Am. B **12**, 2467- (1995).
21. D. N. Fittinghoff, J. L. Bowie, J. N. Sweetser, R. T. Jennings, M. A. Krumbügel, K. W. DeLong, R. Trebino, and I. A. Walmsley, "Measurement of the intensity and phase of ultraweak, ultrashort laser pulses," Opt. Lett. **21**, 884-886 (1996).
22. Pamela Bowlan, Pablo Gabolde, Aparna Shreenath, Kristan McGresham, Rick Trebino, and Selcuk Akturk, "Crossed-beam spectral interferometry: a simple, high-spectral-resolution method for completely characterizing complex ultrashort pulses in real time," Opt. Express **14**, 11892-11900 (2006)
23. http://www.andor.com/scientific_cameras/idus-ingaas/models/?iProductCodeID=71.
24. C. -B. Huang, S. -G. Park, D. E. Leaird, and A. M. Weiner, "Nonlinearly broadened phase-modulated continuous-wave laser frequency combs characterized using DPSK decoding," Opt. Express **16**, 2520-2527 (2008).
25. http://www.home.agilent.com/agilent/product.jspx?pn=N4901B&NEWCCLC=INeng
26. M. Shirasaki. "Large angular dispersion by a virtually imaged phased array and it's application to a wavelength division multiplexer," Opt. Letters. **21**, 366–368, (1996).
27. S. Xiao and A. M. Weiner. "2-D wavelength demultiplexer with potential for >= 1000 channels in the C-band," Optics Express **12**, 2895-2902, (2004).
28. S. A. Diddams, L. Hollberg, and V. Mbele. "Molecular fingerprinting with the resolved modes of a femtosecond laser frequency comb," Nature **445**, 627-630 (2007).
29. Franklyn Quinlan, Charles Williams, Sarper Ozharar, Sangyoun Gee, and Peter J. Delfyett, "Self-Stabilization of the Optical Frequencies and the Pulse Repetition Rate in a Coupled Optoelectronic Oscillator," J. Lightwave Technol. **26**, 2571-2577 (2008).
30. M. Kourogi, K. Nakagawa and M. Ohtsu "Wide-span optical frequency comb generator for accurate optical frequency difference measurement," IEEE J. Quantum Electron. **29**, 2693, (1993).
31. Z. Jiang, D. Leaird, C. B. Huang, H. Miao, M. Kourogi, K. Imai, and A. M. Weiner, "Spectral line-by-line pulse shaping on an optical frequency comb generator," IEEE J. Quantum Electron. **43**, 1163-1174 (2007).
32. M. S. Kirchner, D. A. Braje, T. M. Fortier, A. M. Weiner, L. Hollberg, and S. A. Diddams, "Generation of 20 GHz, sub-40 fs pulses at 960 nm via repetition-rate multiplication," Opt. Lett. **34**, 872-874 (2009).
33. Jian Chen, Jason W. Sickler, Peter Fendel, Erich P. Ippen, Franz X. Kärtner, Tobias Wilken, Ronald Holzwarth, and Theodor W. Hänsch, "Generation of low-timing-jitter femtosecond pulse trains with 2 GHz repetition rate via external repetition rate multiplication," Opt. Lett. **33**, 959-961 (2008).
34. A. Bartels, R. Gebs, M. S. Kirchner, and S. A. Diddams, "Spectrally resolved optical frequency comb from a self-referenced 5 GHz femtosecond laser," Opt. Lett. **32**, 2553-2555 (2007).
35. Li-Jin Chen, Andrew J. Benedick, Jonathan R. Birge, Michelle Y. Sander, and Franz Kärtner, "Octave-spanning, dual-output 2.166 GHz Ti:sapphire laser," Opt. Express **16**, 20699-20705 (2008).

## 1. Introduction

Recently there has been significant activity in optical arbitrary waveform generation (OAWG) in which the amplitude and phase of individual lines of relatively high repetition rate frequency combs are controlled [1-6]. In the time domain, this leads to generation of wide temporal window (up to 100% duty factor) waveforms repeating at the repetition rate of the comb. A schematic of this is shown in the spectral and temporal domains in Figs. 1(a) and 1(b). Also, depending on the total available bandwidth, the temporal features can be made very fine, allowing as a whole, generation of very high complexity optical waveforms. However we can envision an even more interesting regime of operation. By changing the pulse-shaping function at the repetition rate of the comb, potentially infinite record length waveforms with arbitrary temporal resolution can be generated. In this case every successive pulse of the pulse train constituting the frequency comb has a different shape as schematically shown in Fig. 1(c). Such waveforms can significantly benefit areas such as optical communications where one can envision simultaneous encoding in coherent formats of all channels of a dense wavelength division multiplexing (DWDM) system simultaneously; in ranging applications like coherent light detection and ranging (coherent LIDAR) [7], allowing long range and high resolution simultaneously, or in applications in spectroscopy and coherent control [8].

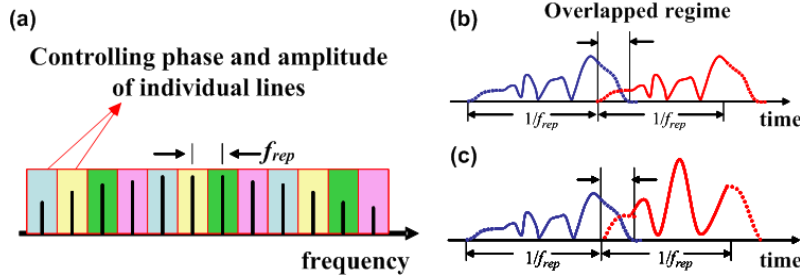

Fig. 1. Schematic of line-by-line shaping, (a) Spectral domain, (b) temporal domain, (c) temporal domain with rapid update.

For utilization of such waveforms, simultaneous to development of generation capabilities, it is also necessary to have suitable waveform characterization apparata. Also, the waveform measurement apparatus by itself will be useful for spectroscopic applications or as detectors similar in essence to streak cameras. Transient behavior accompanying rapid update of waveforms can also be studied [9].

Measuring such waveforms is demanding for several reasons. Firstly, owing to the line-by-line shaped nature of such waveforms, abrupt phase and amplitude changes can occur from line-to-line requiring high spectral resolution together with the ability to handle wide spectral bandwidths. Conventional ultrafast measurement methods which work very well for wide bandwidth (short pulses) with smoothly varying spectra and phase do not work well in this regime [10-12]. Secondly, because of the availability of only a single frame per measurement, very high sensitivity is necessary. Due to high repetition rates of the sources used, for a given average power, there is significantly less energy per pulse (for a 10GHz source, the energy per pulse is 100dB less than the average power) increasing the sensitivity requirement for a single-shot measurement. Thirdly, since acquisition times (limited by available cameras) are significantly slower than the repetition rate of the source, a high quality time-gating system is necessary to select the waveform frame of interest and suppress all the other frames with sufficient fidelity that they do not affect the measurement.

Previously, we adapted a zero-delay version of spectral interferometry to measure line-by-line shaped waveforms using average low power [13] (examples of other efforts for similar characterization applications can be seen in [3], [14]-[15]). Here we build on our previous work to demonstrate for the first time single-shot amplitude and phase characterization of waveforms with ~1 THz of bandwidth from a 10 GHz frequency comb. Along with the need for sufficient sensitivity, in our case of a high repetition rate signal, true single-shot operation also requires a high-quality time gating system to select a single waveform frame. In our experiments we implement both the high extinction time gating system and an apparatus to create test waveforms which are updated at the repetition rate of the source. The latter is necessary to fully probe the capability of the time gating system. In contrast, in the case of static waveforms, suboptimal gating which does not sufficiently extinguish power from other waveform periods will not significantly affect the measurement since different waveform periods are identical.

Finally, it is important to contrast this technique with time domain based techniques [16-18] and repetitive sampling techniques [19]. Although time domain methods for single shot waveform capture have been demonstrated, these techniques acquire only the temporal intensity profile and do not provide phase information. Sampling techniques can be used to measure repetitive waveforms or to acquire eye diagrams or signal constellations for modulated data in telecommunication systems. However, single rare events cannot be captured via sampling approaches and require single shot waveform characterization. In our experiment we achieve complete characterization of both spectral amplitude and phase in a single-shot. We believe that this constitutes a significant advance to the suite of optical arbitrary waveform measurement techniques that have been vigorously investigated in recent work.

## 2. Experimental setup

Spectral interferometry [20] is a well known pulse characterization technique which measures an unknown signal waveform with respect to a characterized reference pulse by looking at the spectrally resolved interference between them. Owing to its linear nature, it adapts well to low power applications. The fact that it is not a self-referenced technique, and needs a well characterized reference pulse is not a significant limitation since a short pulse used as reference can be characterized by other well established self referenced techniques (see [21] for example). However, in conventional implementations of spectral interferometry, in order to unambiguously retrieve the phase information from one component of the interference signal (either in phase or quadrature), a large delay is necessary between the signal pulse and the reference pulse. This leads to very high demands on spectral resolution (many times more than the spectral features in the signal waveform) particularly for optical arbitrary waveforms which already have fine spectral content. In order to minimize spectral resolution requirements, we adapt a version of spectral interferometry called dual-quadrature spectral interferometry which uses polarization demultiplexing to measure the complete interference signal (both in-phase and quadrature) allowing zero-delay operation and hence minimizing spectral resolution requirements. (For another zero-delay method using a two dimensional geometry, see [22]).

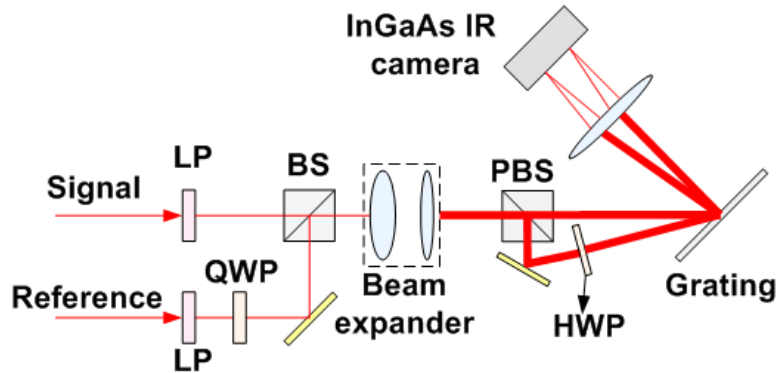

Fig. 2. Setup, LP – linear polarizer aligned at 45deg, QWP – quarter wave plate, BS – beam splitter, PBS – polarizing beam splitter, HWP – half wave plate.

Figure 2 shows the dual-quadrature spectral interferometry setup. The signal to be measured is linearly polarized at a 45 deg angle while the reference is circularly polarized. These two beams are combined followed by a high resolution spectrometer consisting of a 10X beam expander, a 1100line/mm grating, and an InGaAs IR camera [Andor, 23]. The pixel dimension of the camera in the dispersion direction is 25microns, and there are 512 pixels. The spectrometer resolution is 5 GHz per pixel which gives a line-to-line spacing of 2 pixels on the camera (corresponding to the frequency comb spacing of 10 GHz). The spectrometer simultaneously measures the interferograms in both polarizations

(corresponding to the in-phase and quadrature terms) by mapping them to different physical locations on the camera. This is very important for single shot operation since only one waveform period is available per measurement. For both channels the measured spectrometer crosstalk between adjacent comb lines is ~5%. For each of the polarizations, an input bandwidth of 1THz spreads across 200 pixels on the camera. This leaves some freedom to increase the measurement bandwidth (temporal resolution) if necessary. A point to note here is that there is no fundamental limitation on how much signal bandwidth can be handled. By choosing a camera with more pixels, the temporal resolution can be significantly increased depending on the requirements of the application. We retrieve waveform information unambiguously from a single frame of camera data with 1.4μs integration time, which defines our data acquisition time. A mathematical description and retrieval information is discussed in [13].

Before we go further we will briefly describe the source used in our experiments. A detailed account can be found in [24]. The frequency comb source we use is generated by sending a continuous wave (CW) laser through a strongly driven phase modulator followed by an intensity modulator. At this point, though a frequency comb is generated, the temporal envelope is still wide owing to phase variations between different lines. This is corrected using a line-by-line pulse shaper generating bandwidth limited pulses of ~2.5 ps duration (usable bandwidth of ~300GHz). The bandwidth limited nature of the pulse is supported by a close match between the simulated autocorrelation assuming flat spectral phase and the measured autocorrelation [24]. This has also been independently verified with self-referenced spectral shearing interferometry [15]. In some of our experiments, these pulses are then spectrally broadened and compressed using a soliton based dispersion decreasing fiber which generates ~500 fs pulses. The desired bandwidth (~1 THz in these experiments) is selected using a simple pulse shaper based filter. A fraction of the power is used as the reference pulse; the remaining power is used for signal waveform synthesis. The bandwidth limited nature of the reference pulse is again tested by comparing its autocorrelation with the simulated autocorrelation assuming flat spectral phase. This allows us to take the phase of the reference pulse as flat, with the result that the retrieved phase from the spectral interferometry should correspond reasonably closely to the true phase of the signal waveform.

The integration time of our camera is 1.4 μs while the repetition rate of the comb corresponds to 100 ps. When the waveforms are updated rapidly, to make a high quality measurement it becomes necessary to sufficiently suppress all other waveform periods in the integration window other than the period of interest. Fig. 3(a) shows a cartoon depicting this. The extinction requirement is dictated by the integration time of the available camera technology. Since the factor between the integration time and a single waveform frame (i.e., the comb period) is a factor of 14000 (or ~42dB) in our experiments, even if 1/14000 of the power leaks through during every waveform period, it will still integrate up to reduce the contrast between the waveform to be measured and the leakage to ~1. Since a high contrast is desirable to make clean measurements we achieve this by using a cascaded dual intensity modulator scheme (Fig. 3(b)). Here we first use a high extinction ratio modulator (>45dB extinction) followed by a conventional telecommunications modulator with >20dB extinction. The series extinction ratio is >65 dB, which corresponds to a signal to leakage contrast of >100. Both the intensity modulators are driven by an Agilent 13.5Gbps BERT based pattern generator which produces a '1' for 100ps (limited by rise and fall times) [25] and '0's for the remaining part of the 1.4μs window. We note that in our cascaded modulator scheme, the second telecommunications modulator not only provides the extra 20 dB of extinction but also cleans up the rising and falling edges of the gating window. Fig. 3(c) shows the spectrum of a single pulse gated from a 10 GHz frequency comb. The spectrum is taken through an OSA with a resolution of 1.25GHz. What was initially a spectrum made of sharp discrete lines spaced by 10 GHz is now a smooth spectrum with no sign of residual discrete line structure. This is one signature of high quality single pulse gating.

Ideally, if the acquisition time of the camera (which gives the response time of the spectrometer) is the same as the repetition rate of the comb, we would have a continuous time acquisition system without the need for gating (the simple retrieval algorithm from the interferogram data allows us to assume the computation also to be real time). Such a system can measure potentially infinite record length waveforms without any dead spacing. Current camera technology though is still far away from the GHz class acquisition times necessary. However, if required by the application, it is possible to increase the record length per acquisition. Since this depends on the resolution of the spectrometer, by using high resolution spectral dispersers like the virtually imaged phase array (VIPA)[26-28, 6] possibly in conjunction with other dispersers (to obtain simultaneous high resolution – broad bandwidth operation) sub GHz resolution can be obtained, which in turn corresponds to temporal record lengths of >1ns. Temporal demultiplexing can be used to further increase the record lengths. We expect that in future, any candidate for continuous operation would probably involve all these aspects simultaneously to push towards its objective. The longer record lengths per acquisition afforded by the higher spectral resolution and temporal demultiplexing allows for relatively slower cameras to be used in the spectrometer.

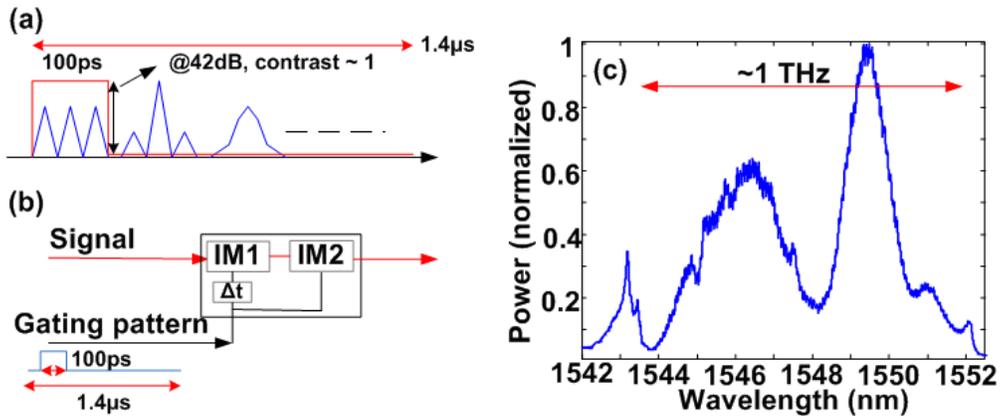

Fig. 3. (a) Schematic showing the need for high extinction gating, (b) Gating scheme, (c) Gated spectrum of a 10GHz pulse train.

With respect to the choice of waveforms for measurement, an OAWG encoder which can generate arbitrary waveforms over a large number of spectral lines and update them at the repetition rate of the comb would be ideal. However, until such technology is available, other simple schemes can be used to generate adequate test waveforms. Some desirable attributes are: (i) abrupt or sharp variations of amplitude and phase in the spectral domain (since ours is a spectral domain based technique), and (ii) fast update since we saw previously that a true single shot operation can only be verified with dynamic waveforms. Our scheme to achieve this is shown in Fig. 4(a). An input 10GHz pulse train is split into two arms; in one arm an intensity modulator removes every alternate pulse. The relative heights of the pulses can also be controlled using attenuators present in each arm. These pulses are then combined with a delay to form a quasi-dynamic signal. Fig. 4(b) shows the sampling scope trace taken using a 60 GHz photodiode showing alternate periods of single pulses and pulse pairs. Though in the time domain they look relatively simple, in the frequency domain these waveforms have rapid amplitude fringes characteristic of two temporally separated pulses interfering with each other and a linear spectral phase with abrupt 0-$\pi$ jumps whenever the amplitude of the interference signal changes sign. Another motivation to choose such a waveform pattern is that, when we retrieve the single pulse waveform, if the gating is not ideal, leakage from adjacent periods is expected to show up as a small satellite pulse at the position of the 2[nd] pulse of the pulse pair waveform. Absence of this can be interpreted as sign of high qualtiy single waveform gating.

Figs. 4(c) and 4(d) respectively show sampling scope traces taken of a pulse pair and of a single pulse waveform when the gating is switched on. By varying the delay of the gating pattern from the pattern generator, different waveform frames can be selected. The gated waveforms are amplified using an erbium doped fiber amplifier (EDFA) prior to measurement.

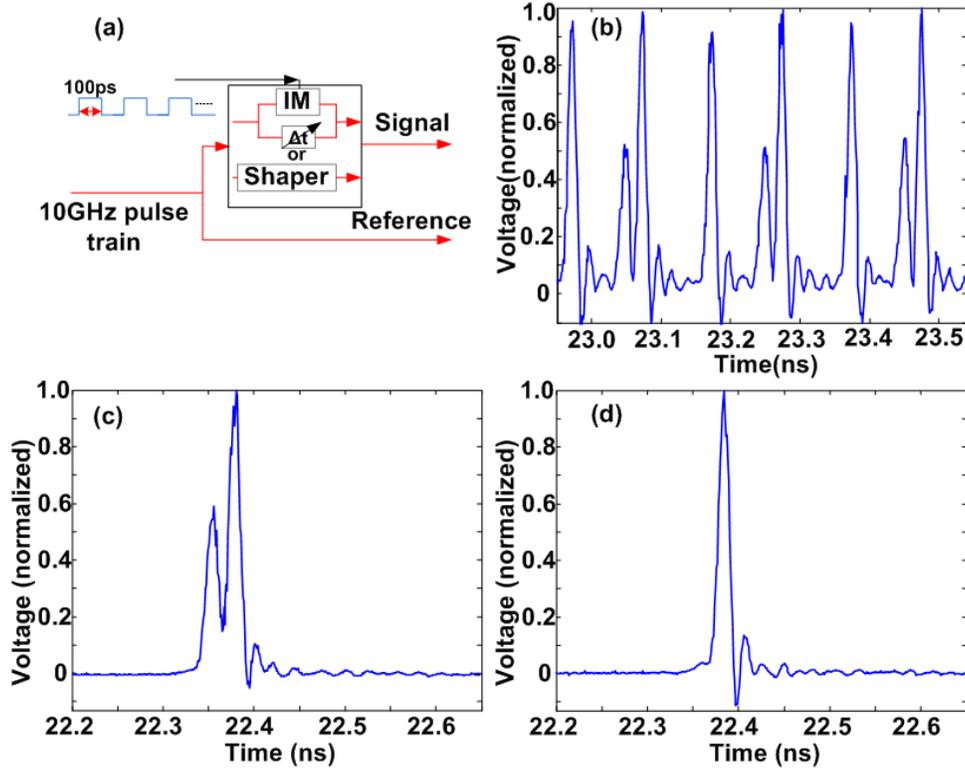

Fig. 4. (a) Scheme for waveform generation, (b) 60 GHz sampling scope trace of waveform without gating, (c) Gated pulse pair, (d) Gated single pulse

In addition to dynamic waveform measurements confirming true single shot operation as described above, we have also performed experiments where the dynamic pulse pair generator is replaced with a static line-by-line pulse shaper [2] (Fig. 4(a) bottom part). In this case we can program the pulse shaper to generate more complex text waveforms which are still measured in single-shot operation but without waveform update. Since waveform update is more a test of the gating system than of the spectral interferometry setup, once the gating system is verified we are free to measure more complex waveforms in static situations.

## 3. Results

Figure 5 shows the experimental results for dynamic waveforms. Fig. 5(a) shows the measured spectral amplitude and phase for a gated single pulse. The phase is relatively flat as expected for a bandwidth limited pulse. As expected the retrieved spectrum resembles the spectrum for the gated pulse as shown in Fig. 3(c) (but flipped because it is plotted in frequency). The spectrum is relatively noisy and this we believe is due to amplified spontaneous emission (ASE) noise added by the amplifier before the spectral broadening process and by the EDFA which amplifies the waveforms after gating. However as far as the measurement is concerned, this is the spectrum of the source in that waveform period. Also we see that the phase plot deviates slightly from its flat nature at places where the signal spectrum is small; this is likely due to reduced signal to noise ration (SNR) at these points causing extra phase errors. Perhaps a better way to look at the data is to calculate the time domain waveform using the retrieved spectrum and phase. This is shown in Fig. 5(b). A clean pulse is seen as expected. Also, no satellite pulse is seen which, as discussed earlier, demonstrates high qualtiy single waveform gating and unambiguous single-shot measurement.

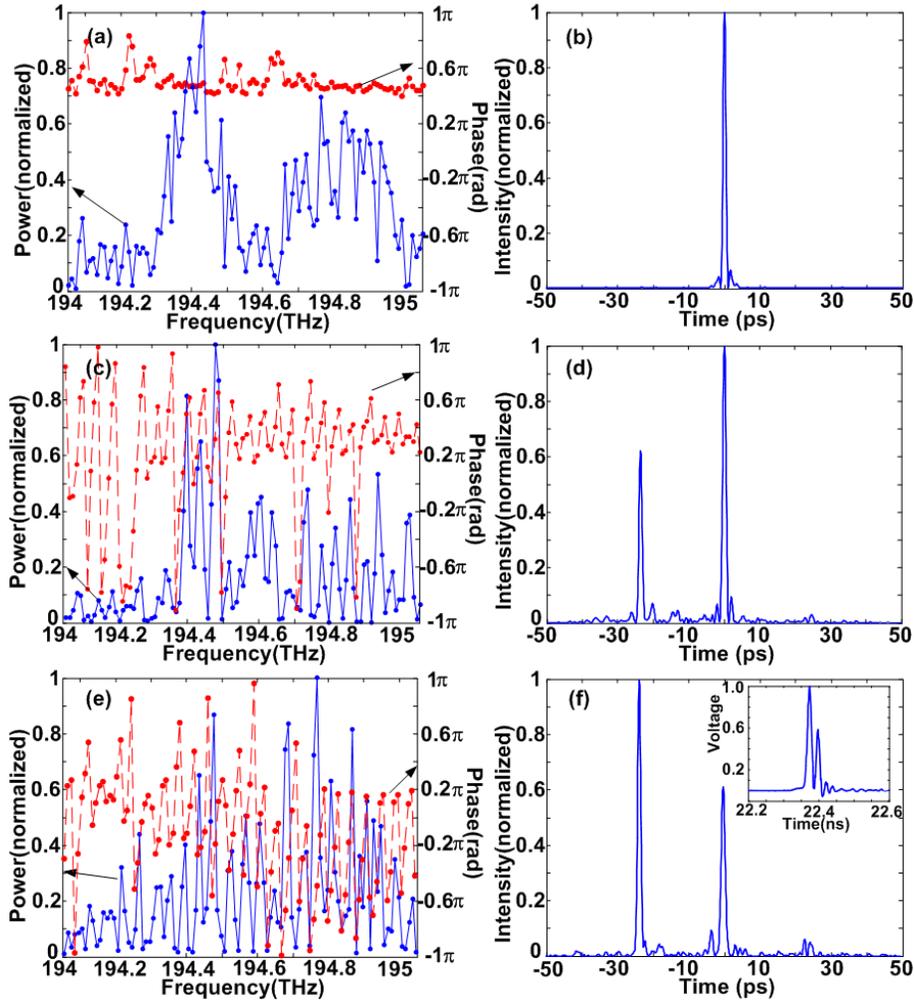

Fig. 5. (a),(b) Retrieved spectrum, phase and generated time domain trace for gated single pulse, (c), (d) For a gated pulse pair, (e), (f) For a gated pulse pair with different relative heights (inset – RF scope trace)

Figures 5(c) and 5(d) show the retrieved spectrum and phase and the generated time domain waveform for the gated pulse pair whose RF scope trace was shown in Fig. 4(c). In the spectral domain we see fast variations both in spectral amplitude and phase - no easily discernible pattern exists, but when the time domain waveform is calculated, we see that it agrees very well with what was expected. This gives strong evidence of proper phase retrieval. Another point to note is that in our experiments the pulse at t=0 is left unchanged, while the pulse at t ~ -27ps is modulated. This point is exactly consistent with our waveform retrieval data, which gives additional evidence of correct phase measurement (a delay in time causes a linear spectral phase). Figs. 5(e) and 5(f) shows the retrieved spectrum and phase and the calculated time domain waveform for a different gated pulse pair waveform where the amplitude of the earlier, gated pulse is increased relative to the ungated pulse at t=0. The sampling scope trace is shown as the inset in Fig. 5(f). Excellent waveform retrieval is observed as indicated by the agreement between the calculated time domain waveform and the scope trace.

After the waveform generation and gating, the energy in a single gated pulse pair waveform was ~5pJ (for a gated single pulse it was roughly half of that). The spectrometer loss in our setup was around 10 dB which leads to about 500 fJ of signal energy distributed over 400 pixels on the camera (200 pixels per channel). The reference pulse energy was chosen such that the spectral intensity of the reference pulse was at least twice as strong as that of the signal pulse. This condition is necessary for unambiguous retrieval [13]. The efficiency of the camera was around 70% which corresponds to an average of ~6700 photoelectrons from the signal waveform per pixel. The specified noise for the camera is ~600 photoelectrons per pixel. Since this is an interferometric measurement, depending on the coherent sum of the reference waveform and the signal waveform at each pixel (which depends on the phase difference) the SNR varies for different pixels. In cases where the sum is low, the SNR is lower leading to reduced accuracy of measurement. On average we observed an SNR of around 5 (20% contribution by noise) which corresponds to a higher noise contribution than by just considering camera noise. This we believe is due to amplifier ASE noise.

Fig. 6 shows measurement results with user defined waveforms generated using a line-by-line pulse shaper. The bandwidth in these experiments is around 200GHz limited largely by the bandwidth available in the frequency comb before spectral broadening. Fig. 6(a) shows the retrieved spectral phase (circles) and the applied phase (a quadratic, shown as solid line). Excellent agreement is observed. Fig. 6(b) show the retrieved phase when cubic spectral phase is applied. The errors (standard deviation of differences between applied and retrieved phase) are $0.11\pi$ and $0.13\pi$, respectively. These are only slightly higher than the errors of approximately $0.1\pi$ observed in previous measurements at higher average power in which spectral interferometry data were acquired for static waveforms over multiple waveform periods by integrating over the 1.4 μsec camera integration time (corresponds to $>10^4$ waveform periods) [13].

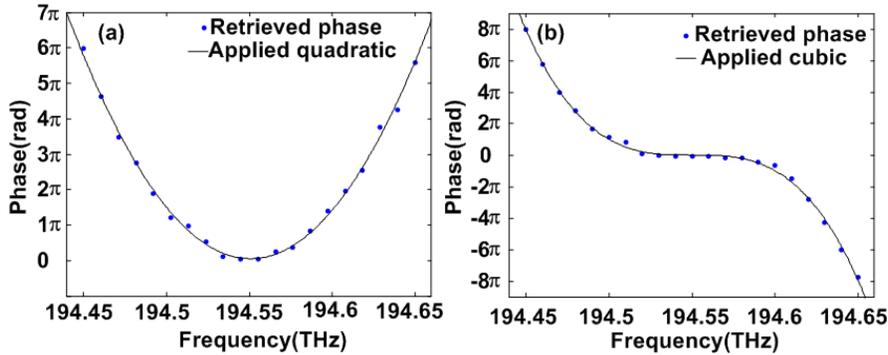

Fig. 6. (a) Applied quadratic phase and retrieved phase, (b) Applied cubic phase and retrieved phase.

## 5. Summary and future work

In summary, we have used time gated dual-quadrature spectral interferometry to demonstrate for the first time unambiguous single-shot characterization of both spectral amplitude and phase of optical arbitrary waveforms generated by line-by-line pulse shaping of an optical frequency comb. Our experiments accommodate arbitrary waveforms with up to 1 THz optical bandwidth (corresponding to a temporal resolution of 1 ps) and a spectral line spacing of 10 GHz (corresponding to a record length of 100 ps). The sensitivity of our approach together with the ability to measure high complexity optical waveforms offers potential for impact in spectroscopic applications, in communications, and in observation of fast transient phenomena in a variety of fields.

Although our current experiments have been performed with a comb generated by direct modulation of a continuous-wave laser, as is becoming increasingly common in telecommunications, our approach is applicable to measurement of optical arbitrary waveforms generated from a variety of other high repetition rate comb sources. These include short pulse coupled optoelectronic oscillators [29], optical frequency comb generators based on synchronous modulation in an optical cavity [30, 31], mode-locked lasers externally filtered to obtain high repetition rate [32, 33], and self-referenced lasers mode-locked directly at high repetition rate [34, 35]. Furthermore, there is no fundamental limitation on the optical bandwidth that can be accommodated. From a practical perspective, generalizing our approach to use two-dimensional disperser geometries [27, 6] compatible with camera technologies comprising hundreds of thousands of detector elements should allow optical arbitrary waveform characterization even for optical arbitrary waveforms with bandwidths approaching the octave regime.

**Acknowledgements**

This work is supported by DARPA/ARO under grant W911NF-07-1-0625 as part of the DARPA Optical Arbitrary Waveform Generation (OAWG) program and by the National Science Foundation under grant ECCS-0601692.